\documentclass[aps,pra,twocolumn,groupedaddress]{revtex4-1}

\usepackage{color}
\usepackage{graphicx}				
\usepackage{soul}

\usepackage{natbib}

\begin{document}

\title{A Nonlinear Schr\"{o}dinger Wave Equation With Linear Quantum Behavior}
\author{Chris D. Richardson}
\author{Peter Schlagheck}
\author{John Martin}
\author{Nicolas Vandewalle}
\author{Thierry Bastin}
\affiliation{D\'epartement de Physique, University of Liege, 4000 Liege, Belgium}
\date{\today}							

\begin{abstract}

We show that a nonlinear Schr\"{o}dinger wave equation can reproduce all the features of linear quantum mechanics.  This nonlinear wave equation is obtained by exploring, in a uniform language,  the transition from fully classical theory governed by a nonlinear classical wave equation to quantum theory.  The classical wave equation includes a nonlinear classicality enforcing potential which when eliminated transforms the wave equation into the linear Schr\"{o}dinger equation.  We show that it is not necessary to completely cancel this nonlinearity to recover the linear behavior of quantum mechanics.  Scaling the classicality enforcing potential is sufficient to have quantum-like features appear and is equivalent to scaling Planck's constant.

\end{abstract}

\maketitle

\section{Introduction}

The boundary between where quantum mechanics ends and classical mechanics begins is being pushed by experiments operating far from the microscopic scale in which quantum behavior is normally associated.  For example, Arndt \emph{et al.}~\cite{bib:arndt} observed interference between mesoscopic fullerenes and Lee \emph{et al.}'s~\cite{bib:leediamonds} experiment entangled the vibrational modes of two macroscopic diamonds.  It is therefore important to be able to describe the manner in which quantum theory transitions into classical theory and to be able to define what fundamental elements are responsible for the separation between the two theories.  There is no doubt that all the behavior of classical mechanics is contained entirely within quantum theory but it is also obvious that the two regimes must, for practical reasons, be treated differently.

One of the main differences between the two theories is the linearity of the governing equations.  The linearity inherent in quantum mechanics is evident from the Schr\"{o}dinger wave equation and has been experimentally verified~\cite{PhysRevLett.65.2931} to extreme accuracy.  In contrast, it is well known that classical mechanics is intrinsically nonlinear.  When expressed in the language of quantum mechanics, classical mechanics is governed by a nonlinear wave equation~\cite{bib:obm} similar in form to the Schr\"{o}dinger equation but which allows no quantum or wavelike features.  The switch from classical theory to quantum theory and to the Schr\"{o}dinger equation can then be brought about by eliminating the nonlinearity from the classical wave equation~\cite{bib:revisited}, for quantum mechanics is after all a linear theory.

Here we show that there is a subtle relationship between the linear character of quantum mechanics and the linearity of the governing wave equation.  Linear behavior which we associate with quantum mechanics, like superposition or interference, resulting from the linear Schr\"{o}dinger equation can also come from a wave equation which is nonlinear, a fact which may impact branches of physics reliant on nonlinear governing equations such as hydrodynamics and condensed matter physics.  In this paper we construct this wave equation and name it the \emph{transition} equation as it can be tuned to describe both classical and quantum behavior (Sec.~\ref{sec:transitionequation}).  We demonstrate the quantum-like behavior of this transition equation by revealing its equivalence to the linear Schr\"{o}dinger equation with a rescaled Planck's constant (Sec.~\ref{sec:scaling}).  We then reinforce this equivalence by numerically exploring the standard single particle interference problem (Sec.~\ref{sec:inter}). Finally, we draw conclusions in Sec.~\ref{sec:conclusion}.
\section{The Transition Equation\label{sec:transitionequation}}

Despite classical non-relativistic mechanics being completely contained in non-relativistic quantum mechanics, the two regimes are discussed in quite different languages.  To develop a tool that smoothly transitions between the two regimes we must be able to describe them both using the same language.  Wave functions, wave equations, and probability densities are the language of quantum mechanics which is governed by the Schr\"{o}dinger wave equation,
\begin{eqnarray}
i \hbar \frac{\partial \psi(\mathbf{r},t)}{\partial t} = - \frac{\hbar^2}{2 m} \nabla^2 \psi(\mathbf{r},t) + V(\mathbf{r},t) \psi(\mathbf{r},t) \;, \label{eqn:schrod}
\end{eqnarray}
where $\psi(\mathbf{r},t)$ is a wave function whose modulus squared gives the probability density of finding a particle at position $\mathbf{r}$ and time $t$, $m$ is the mass of the particle, $V$ is the potential experienced by the particle, and $\hbar$ is the reduced Planck's constant.  Classical mechanics on the other hand uses the language of trajectories and is governed by Newton's laws.  We can formulate classical mechanics so that it can be expressed entirely by the Hamilton-Jacobi~\cite{bib:hamiltonjacobi} equation,
\begin{eqnarray}
\frac{\partial S_{c}(\mathbf{r},t)}{\partial t} &=& - \frac{1}{2 m} \left[\nabla S_{c}(\mathbf{r},t)\right]^2 - V(\mathbf{r},t) \;, \label{eqn:hamjac}
\end{eqnarray}
where $S_{c}$ is the classical action which defines a canonical transformation between initial and final phase-space coordinates.  For a given initial condition the behavior of a classical particle will be described by a trajectory derived wholly~\cite{bib:hamiltonjacobi} from the action via $m \dot{\mathbf{r}} = \nabla S_{c}$.  

The description of quantum mechanics using classical language or the hydrodynamic form of quantum mechanics has been well known since Madelung~\cite{bib:madelung} in 1926 and was revived by Bohm~\cite{bib:bohm} in 1952.  It begins by expressing the complex-valued wave function in its polar form,
\begin{eqnarray}
\psi(\mathbf{r},t) = A(\mathbf{r},t) e^{ i S(\mathbf{r},t) / \hbar } \;, \label{eqn:polarwf}
\end{eqnarray}
where $A$ is the real-valued amplitude and $S/\hbar$ is the real-valued phase.  Plugging this into the Schr\"{o}dinger equation, Eq.~(\ref{eqn:schrod}), we obtain two equations.   The first is the continuity equation which expresses the conservation of probability,
\begin{eqnarray}
\frac{\partial A(\mathbf{r},t)}{\partial t} &=& - \frac{1}{m} \left[\nabla A(\mathbf{r},t)\right] \cdot [\nabla S(\mathbf{r},t)]  \label{eqn:continuity} \\
&& - \frac{1}{2 m} A(\mathbf{r},t) \nabla^2 S(\mathbf{r},t) \;, \nonumber
\end{eqnarray}
which is equivalent to $\frac{\partial \rho}{\partial t} + \nabla \cdot \mathbf{j} = 0$, where $\rho = \left| \psi \right|^2=  A^2$ and $\mathbf{j} = \frac{\hbar}{2 i m} (\psi^* \nabla \psi - \psi \nabla \psi^*) = \frac{1}{m }A^2 \nabla S$ is the probability density current. The second is a Hamilton-Jacobi-like equation:
\begin{eqnarray}
\frac{\partial S(\mathbf{r},t)}{\partial t} &=& - \frac{1}{2 m} \left[\nabla S(\mathbf{r},t)\right]^2 - \left[ V(\mathbf{r},t) + U(\mathbf{r},t) \right] \;. \label{eqn:hamjacbohm}
\end{eqnarray}
This equation differs in form from the Hamilton-Jacobi equation, Eq.~(\ref{eqn:hamjac}), by an extra potential,
 \begin{eqnarray}
U(\mathbf{r},t) &=& - \frac{\hbar^2}{2 m} \frac{ \nabla^2 A(\mathbf{r},t)}{A(\mathbf{r},t)} = - \frac{\hbar^2}{2 m} \frac{\nabla^2 \left| \psi(\mathbf{r},t) \right|}{\left| \psi(\mathbf{r},t) \right|}  \label{eqn:qmp}
\end{eqnarray}
which Bohm called the \emph{quantum-mechanical potential}.  Adding this potential to the classical Hamilton-Jacobi, Eq.~(\ref{eqn:hamjac}), allows us to derive an action which in turn gives a trajectory.  However, unlike a classical trajectory this one can have quantum behavior and non-Newtonian motion.

The description of classical mechanics using quantum language is less well known but has been derived recently by Oriols and Mompart~\cite{bib:obm}. If a particle's initial position is only known through a probability distribution, $A_c^2(\mathbf{r},0) \,\mathrm{d}^3 r$, the classical trajectories resulting from each of these initial positions $\mathbf{r}$ will evolve according to the Hamilton-Jacobi equation and give at any time $t>0$ the probability distribution $A_c^2(\mathbf{r},t) \,\mathrm{d}^3 r$. A classical wave function similar to Eq.~(\ref{eqn:polarwf}), $\psi_{c}(\mathbf{r},t) = A_{c}(\mathbf{r},t) \exp\left[i S_{c}(\mathbf{r},t) / \hbar\right]$, can be constructed where $\hbar$ is used to provide a dimensionless argument and $S_{c}$ is again the classical action from the Hamilton-Jacobi equation, Eq.~(\ref{eqn:hamjac}).  Using this form of the wave function Oriols and Mompart~\cite{bib:obm} derive a wave equation, similar in form to the Schr\"{o}dinger equation, that describes the evolution of a classical particle.  We call it the classical Schr\"{o}dinger-like equation and it is given by
\begin{eqnarray}
i \hbar \frac{\partial \psi_{c}(\mathbf{r},t)}{\partial t} &=& - \frac{\hbar^2}{2 m} \nabla^2 \psi_{c}(\mathbf{r},t) + V(\mathbf{r},t) \psi_{c}(\mathbf{r},t) \label{eqn:class_schrod} \\ 
&&+ \frac{\hbar^2}{2 m} \frac{\nabla^2 \left| \psi_{c}(\mathbf{r},t) \right|}{\left| \psi_{c}(\mathbf{r},t) \right|} \psi_{c}(\mathbf{r},t) \;, \nonumber 
\end{eqnarray}
where the probability density is given from the modulus squared of the wave function, $\rho_c = A_c^2 = \left|\psi_{c}\right|^2$, in analogy to quantum mechanics.  Equation~(\ref{eqn:class_schrod}) while having completely classical behavior is similar in form to the Schr\"{o}dinger equation except for an extra nonlinear term which has the effect of canceling out all quantum and wave-like effects.  It is of course Bohm's quantum-mechanical potential, Eq.~(\ref{eqn:qmp}), with the opposite sign, $-U(\mathbf{r},t)$.

While exploring the origin of the Schr\"{o}dinger equation Schleich \emph{et al.}~\cite{bib:revisited} also derive the nonlinear classical Schr\"{o}dinger-like equation, Eq.~(\ref{eqn:class_schrod}), and label the nonlinear term the \emph{classicality-enforcing potential}.  They transfer from Eq.~(\ref{eqn:class_schrod}) to the linear Schr\"{o}dinger equation, Eq.~(\ref{eqn:schrod}), by first making the ansatz $\psi_{c} \equiv \psi$.   They then define a \emph{quantum action} which includes the classicality-enforcing potential.  This leads to the cancellation of the classicality-enforcing potential in Eq.~(\ref{eqn:class_schrod}) and recovery of the Schr\"{o}dinger equation and quantum mechanics.  Schleich \emph{et al.}~\cite{bib:revisited} claim that to recover quantum mechanics Eq.~(\ref{eqn:class_schrod}) must become linear by the complete elimination of the classicality-enforcing potential.  We find that by scaling and not necessarily eliminating the classicality-enforcing potential we can reproduce quantum behavior and recover the linear Schr\"{o}dinger equation with a rescaled Planck's constant.  We insert a \emph{degree of quantumness} $\epsilon$, where $0 \leq \epsilon \leq 1$, into Eq.~(\ref{eqn:class_schrod}) which scales the classicality-enforcing potential and gives
\begin{eqnarray}
i \hbar \frac{\partial \psi_\epsilon(\mathbf{r},t)}{\partial t} &=& - \frac{\hbar^2}{2 m} \nabla^2 \psi_\epsilon(\mathbf{r},t) + V(\mathbf{r},t) \psi_\epsilon(\mathbf{r},t) \label{eqn:class_schrod_ep} \\ 
&&+ (1 - \epsilon) \frac{\hbar^2}{2 m} \frac{\nabla^2 \left| \psi_\epsilon(\mathbf{r},t) \right|}{\left| \psi_\epsilon(\mathbf{r},t) \right|} \psi_\epsilon(\mathbf{r},t) \;. \nonumber
\end{eqnarray}
We call this the transition equation.  For $\epsilon = 1$ it is equal to the Schr\"{o}dinger equation and $\psi_\epsilon \equiv \psi$.  For $\epsilon = 0$ it is equal to the nonlinear classical Schr\"{o}dinger-like equation and $\psi_\epsilon \equiv \psi_c$.  For all other values $0 < \epsilon < 1$ we show that this nonlinear equation exhibits quantum behavior despite the continued presence of the classicality-enforcing potential.

\section{Equivalence to Scaling Planck's Constant\label{sec:scaling}}

Indeed, the nonlinear transition equation can be shown to be equivalent to the linear Schr\"{o}dinger equation with Planck's constant scaled by the degree of quantumness according to
\begin{eqnarray}
\tilde{\hbar} = \hbar \sqrt{\epsilon} \;. \label{eqn:hbar_scaled}
\end{eqnarray}
To this end, we define the polar form of the wave function that satisfies the transition equation to be
\begin{eqnarray}
\psi_\epsilon(\mathbf{r},t) &=& A_\epsilon(\mathbf{r},t) e^{ i S_\epsilon(\mathbf{r},t) / \hbar} \;.
\end{eqnarray}
Inserting this into the transition equation, Eq.~(\ref{eqn:class_schrod_ep}), and finding the individual elements gives
\begin{eqnarray}
\nabla^2 \psi_\epsilon(\mathbf{r},t) &=& \Big\{ \nabla^2 A_\epsilon(\mathbf{r},t) + 2 \frac{i}{\hbar}  [\nabla A_\epsilon(\mathbf{r},t)] \cdot [\nabla S_\epsilon(\mathbf{r},t)] \nonumber  \\
&+& \frac{i}{\hbar} A_\epsilon(\mathbf{r},t) \nabla^2 S_\epsilon(\mathbf{r},t) - \frac{1}{\hbar^2}  A_\epsilon(\mathbf{r},t) [\nabla S_\epsilon(\mathbf{r},t)]^2 \Big\} \nonumber \\
&\times& e^{ i S_\epsilon(\mathbf{r},t) / \hbar } \;,  \label{eqn:cont_laplacian}  \\
i \hbar \frac{\partial \psi_\epsilon(\mathbf{r},t)}{\partial t} &=& \left[i \hbar \frac{\partial A_\epsilon(\mathbf{r},t)}{\partial t} - A_\epsilon \frac{\partial S_\epsilon(\mathbf{r},t)}{\partial t}\right] e^{ i S_\epsilon(\mathbf{r},t) / \hbar } \; \label{eqn:cont_polar}
\end{eqnarray}
and
\begin{eqnarray}
 \frac{\nabla^2 \left| \psi_\epsilon(\mathbf{r},t) \right|}{\left| \psi_\epsilon(\mathbf{r},t) \right|}  \psi_\epsilon(\mathbf{r},t) &=&  [\nabla^2 A_\epsilon(\mathbf{r},t)] e^{ i S_\epsilon(\mathbf{r},t) / \hbar } \;, \label{eqn:cont_cp}
\end{eqnarray}
where Eq.~(\ref{eqn:cont_laplacian}) is the Laplacian, Eq.~(\ref{eqn:cont_polar}) is the time derivative of the polar wave function and Eq.~(\ref{eqn:cont_cp}) is the contribution from the classicality-enforcing potential.

Gathering the real and imaginary terms we obtain the continuity equation, Eq.~(\ref{eqn:continuity}) with $A$ and $S$ replaced by $A_\epsilon$ and $S_\epsilon$, and an equation very similar to the Hamilton-Jacobi equation, Eq.~(\ref{eqn:hamjac}).  It differs by the addition of Bohm's quantum-mechanical potential scaled by the degree of quantumness,
\begin{eqnarray}
 \frac{\partial S_\epsilon(\mathbf{r},t)}{\partial t} &=& - \frac{1}{2 m} [\nabla S_\epsilon(\mathbf{r},t)]^2  \\
 &&- \left[V(\mathbf{r},t) - \epsilon \frac{\hbar^2}{2 m} \frac{ \nabla^2 A_\epsilon(\mathbf{r},t)}{A_\epsilon(\mathbf{r},t)}\right] \;. \nonumber
\end{eqnarray}

This equation can be made to have the appearance of the Schr\"{o}dinger equation by making the substitution $\tilde{\hbar} = \hbar \sqrt{\epsilon}$.   Performing this substitution yields
\begin{eqnarray}
\frac{\partial S_\epsilon(\mathbf{r},t)}{\partial t} &=& - \frac{1}{2 m} [\nabla S_\epsilon(\mathbf{r},t)]^2 \label{eqn:hamilton-jacobi_rescaled} \\
&& -\left[V(\mathbf{r},t) - \frac{\tilde{\hbar}^2}{2 m} \frac{ \nabla^2 A_\epsilon(\mathbf{r},t)}{A_\epsilon(\mathbf{r},t)}\right] \;, \nonumber
\end{eqnarray}
by which means the degree of quantumness is removed.  Equation~(\ref{eqn:continuity}) with $A$ and $S$ replaced by $A_\epsilon$ and $S_\epsilon$ and Eq.~(\ref{eqn:hamilton-jacobi_rescaled}) are now completely equivalent to the Schr\"{o}dinger equation with a rescaled $\hbar$ and we can write a scaled Schr\"{o}dinger-like equation and the associated wave function,
\begin{eqnarray}
\tilde{\psi}(\mathbf{r},t) &\equiv& A_\epsilon(\mathbf{r},t) e^{i S_\epsilon(\mathbf{r},t) / \tilde{\hbar}} \label{eqn:tildewf} \\
&=& \psi_\epsilon(\mathbf{r},t) e^{i S_\epsilon(\mathbf{r},t) (1 / \sqrt{\epsilon} - 1)/\hbar} \;, \\
i \tilde{\hbar} \frac{\partial \tilde{\psi}(\mathbf{r},t)}{\partial t}  &=& -  \frac{\tilde{\hbar}^2}{2 m} \nabla^2 \tilde{\psi}(\mathbf{r},t) + V(\mathbf{r},t) \tilde{\psi}(\mathbf{r},t) \;, \label{eqn:scaled_schrod}
\end{eqnarray}
and note that $| \tilde{\psi}(\mathbf{r},t)|^2 = | \psi_\epsilon(\mathbf{r},t)|^2$.  We also note that as $\tilde{\hbar} \rightarrow 0$ the phase of Eq.~(\ref{eqn:tildewf}) will begin to vary rapidly compared to its amplitude.  This is a necessary assumption of the WKB approximation~\cite{bib:wkb} which leads to accurate results in the semiclassical regime away from the classical turning points.

Note that what we have just done does not correspond to a linearization of an intrinsically nonlinear equation insofar as we have performed no linear approximation.  In contrast, recent work performed by Sbitnev~\cite{bib:navstoke}, while leading to an equivalent scaled Schr\"{o}dinger equation, Eq.~(\ref{eqn:scaled_schrod}), involves a linearization of the nonlinear Navier-Stokes equation.  He did this in order to obtain a theoretical model of a bouncing droplet system~\cite{bib:dropsdiffract}, a macroscopic system which mimics linear quantum behavior.   He manipulates the classical Navier-Stokes equation to derive an equation similar in form to the classical  Schr\"{o}dinger-like equation, Eq.~(\ref{eqn:class_schrod}).  He then makes an approximation that removes the nonlinear term, as did Schleich \emph{et al.}~\cite{bib:revisited} with their nonlinear classical wave equation, and ends up with a governing equation equivalent to the Schr\"{o}dinger equation with Planck's constant replaced by a macroscopic equivalent.  This equivalent Planck's constant of Sbitnev's is comparable to Planck's constant scaled by an appropriate degree of quantumness, $\epsilon$, and as such, even though Sbitnev's governing equation is obtained through a linearization, it is comparable to the scaled Schr\"{o}dinger equation, Eq.~(\ref{eqn:scaled_schrod}), and therefore to the transition equation, Eq.~(\ref{eqn:class_schrod_ep}).

\section{The Interference of Two Wave Packets\label{sec:inter}}

To further demonstrate the quantum and linear behavior of the nonlinear transition equation, Eq.~(\ref{eqn:class_schrod_ep}), we solve it numerically in the context of the standard single particle interference problem.  Interference is not necessarily a uniquely quantum phenomenon.  It can happen with any kind of wave.  However, in the limit of the transition equation becoming classical all wave behavior is suppressed.  Therefore, while not being unique to quantum mechanics, interference can be used as a measure of the quantumness between the two regimes in question.  Using quantum mechanics we would expect two wave packets to spread with time and be represented by the standard Young interference pattern.  For classical particles with no wave nature we expect the packets to maintain their shape and not interfere for all time.  We first solve the system using the scaled linear Schr\"{o}dinger equation, Eq.~(\ref{eqn:scaled_schrod}), and then compare the results to numerical solutions of the nonlinear transition equation, Eq.~(\ref{eqn:class_schrod_ep}).  Thus along with the analytic equivalence previously shown we also numerically demonstrate the equivalence of the two equations in this particular context.

\subsection{Scaled quantum wave packet behavior}

We start the scaled quantum analysis with two Gaussians in one dimension with $V = 0$ and the initial condition
\begin{eqnarray}
\tilde{\psi}(x,0) &=& \sqrt{N_0} \Big[ e^{-(x-d)^2/4 \sigma^2} +e^{-(x+d)^2/4 \sigma^2}\Big] \label{eqn:double_init}
\end{eqnarray}
where $d$ is the distance from the origin to the centers of the Gaussians and $\sigma$ is the root mean square (rms) width. The normalization is
\begin{eqnarray}
 N_0 = \left[2 \sqrt{2 \pi } \sigma  \left(e^{-d^2/2 \sigma ^2}+1\right)\right]^{-1} \;.
 \end{eqnarray}
Using this initial condition to solve the scaled one dimensional time-dependent Schr\"{o}dinger equation,
\begin{eqnarray}
i \tilde{\hbar} \frac{\partial \tilde{\psi}}{\partial t} &=& -\frac{\tilde{\hbar}^2}{2 m} \frac{\partial^2 \tilde{\psi}}{\partial x^2} \;,
\end{eqnarray}
the time-dependent wave function is found to be
\begin{eqnarray}
\tilde{\psi}(x,t) &=& \sqrt{\frac{N_0}{\tilde{a}_t}} \left(e^{-(x-d)^2/4 \tilde{a}^2_t}+e^{-(x+d)^2/4 \tilde{a}^2_t}\right).
\end{eqnarray}
where $\tilde{a}^2_t =  \sigma^2 + \frac{1}{2} i \tilde{\hbar} t / m$.  When the modulus is squared the interference term becomes obvious,
\begin{eqnarray}
\left| \tilde{\psi}(x,t) \right|^2 &=& \frac{N_0}{\tilde{\sigma}_t} \Bigg[\left(e^{-(x-d)^2/4 \tilde{\sigma}^2_t}+e^{-(x+d)^2/4 \tilde{\sigma}^2_t}\right)^2 \label{eqn:wf_double_t}  \\
&&- 4 e^{-\left(x^2 + d^2\right)/2 \tilde{\sigma}^2_t} \sin^2 \left(\frac{\tilde{\hbar} t x d}{4 m \sigma^2 \tilde{\sigma}^2_t} \right)\Bigg] \;,\nonumber
\end{eqnarray}
where $\tilde{\sigma}^2_t = \tilde{\hbar}^2 t^2 / \left(4 m^2 \sigma^2\right) +\sigma^2$ is the time dependent rms width.  This is the expected Young interference pattern.

\subsection{Simulation in the framework of the transition equation}

\begin{figure}[ht]
  \includegraphics[width=1\linewidth]{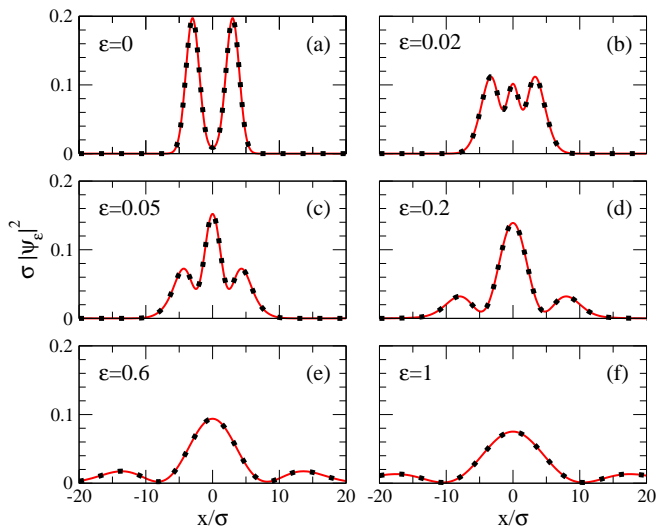}
\caption{(color online)  Interference patterns for various degrees of quantumness, $\epsilon$.  The (red) solid line is the analytic probability density using the Schr\"{o}dinger equation with a scaled $\tilde{\hbar} = \hbar \sqrt{\epsilon}$, Eq.~(\ref{eqn:scaled_schrod}), and the (black) dotted line is the simulated probability density using the transition equation, Eq.~(\ref{eqn:class_schrod_ep}).  All plots are evaluated at the same time $t = 20 m \sigma^2 / \hbar$ and the initial distance from the origin to the center of the two initial Gaussians is $d = 3 \sigma$. Plot (a) is the fully classical case with $\epsilon = 0$. (b) $\epsilon = 0.02$. (c) $\epsilon = 0.05$. (d) $\epsilon = 0.2$. (e) $\epsilon = 0.6$. Plot (f) is the fully quantum case with $\epsilon = 1$ and $\tilde{\hbar} = \hbar$.}
\label{fig:diffract_movie}
\end{figure}

We now numerically solve this standard problem using the nonlinear transition equation, Eq.~(\ref{eqn:class_schrod_ep}), which in one dimension and with $V = 0$ is
\begin{eqnarray}
i \hbar \frac{\partial \psi_\epsilon(x,t)}{\partial t} &=& -\frac{\hbar^2}{2 m} \frac{\partial^2 \psi_\epsilon(x,t)}{\partial x^2}  \label{eqn:schrod_class_nov} \\
&&+ \frac{\hbar^2}{2 m} \frac{1 - \epsilon}{\left| \psi_\epsilon(x,t) \right|} \frac{\partial^2 \left| \psi_\epsilon(x,t) \right|}{\partial x^2} \psi_\epsilon(x,t) \;. \nonumber
\end{eqnarray}
The equation is solved using the explicit finite difference method.  The asymptotic behavior is as expected.  For the case in which the degree of quantumness $\epsilon = 1$ and $\tilde{\hbar} = \hbar$ the interference pattern that forms, Fig.~\ref{fig:diffract_movie}(f), is identical to the standard quantum case in which the transition equation reduces to the Schr\"{o}dinger equation.  For the case in which $\epsilon = 0$ and $\tilde{\hbar} = 0$ the probability density that forms, Fig.~\ref{fig:diffract_movie}(a), is that of the initial distribution, Eq.~(\ref{eqn:double_init}), and is equal to the classical case.  As can be seen in all the frames of Fig.~\ref{fig:diffract_movie} for all values of $\epsilon$ the plots from the numerically solved nonlinear transition equation overlap the plots derived from the scaled Schr\"{o}dinger equation.

Figure~\ref{fig:diffract_movie} demonstrates the equivalence between the nonlinear transition equation, Eq.~(\ref{eqn:class_schrod_ep}), and the linear Schr\"{o}dinger equation with a scaled Planck's constant, Eq.~(\ref{eqn:scaled_schrod}).  For all values of $0 < \epsilon \leq 1$ an interference pattern develops, but the degree of quantumness corresponds to a retardation of the rate in which the interference pattern forms.  Given enough time the pattern will develop into the usual far-field Young interference pattern with a visibility of one for all $0 < \epsilon \leq 1$.  As can be deduced from Fig.~\ref{fig:diffract_movie} the time for a diffraction pattern to develop increases to infinity as the degree of quantumness diminishes.  The only value in which no interference is observed is that for $\epsilon = 0$.  When using the transition equation classical mechanics is a special singular case.

\section{Conclusion\label{sec:conclusion}}

In summary, we have demonstrated both analytically and numerically that it is not necessary to get rid of the classicality-enforcing potential appearing in Oriols and Mompart's~\cite{bib:obm} classical Schr\"{o}dinger-like equation to recover behavior similar to that of quantum mechanics.  We have found that by using a degree of quantumness to scale but not necessarily eliminate the nonlinear classicality-enforcing potential we can construct a transition equation, Eq.~(\ref{eqn:class_schrod_ep}), which we showed to be equivalent to the linear Schr\"{o}dinger equation but with a rescaled Planck's constant, Eq.~(\ref{eqn:scaled_schrod}).

It is interesting that the special behavior observed in the linear theory of quantum mechanics can be reproduced with a nonlinear wave equation such as the transition equation.  This linear quantum behavior is obvious in the transition equation when the degree of quantumness is equal to one; however, when the degree of quantumness is anywhere between zero and one, where nonlinearity is introduced, quantum behavior is still observed.  The nonlinear transition equation mimics the linear Schr\"{o}dinger equation and pure classical mechanics is only observed for the singular case when the degree of quantumness vanishes completely.

In general, it may not be necessary to explicitly eliminate a nonlinearity (e.g., through a linearization procedure) in order to encounter linear behavior, such as wave interference or the superposition principle, in the framework of a nonlinear theory.  This insight might be of interest to other fields of physics that are governed by an intrinsically nonlinear equation, such as fluid dynamics.

\begin{acknowledgments}
This work was financially supported by the Actions de Recherches Concert\'ees (ARC) of the Belgium Wallonia-Brussels Federation under Contract No.~12-17/02.
\end{acknowledgments}

\bibliography{Biblio}
\end{document}